\documentclass[doublecol]{epl2}

\usepackage{amsmath,amssymb,graphicx}

\title{Superscars in the LiNC$\leftrightharpoons$LiCN isomerization reaction}
\shorttitle{Superscars}

\author{S. D. Prado\inst{1,2} \and E. Vergini\inst{3} \and R. M. Benito\inst{3}
  \and F. Borondo\inst{1,4}}
\shortauthor{S. D. Prado \etal}

\institute{
  \inst{1} Departamento de Qu\'{\i}mica C--IX,
           Universidad Aut\'onoma de Madrid, Cantoblanco, 28049 Madrid, Spain\\
  \inst{2} Permanent address: Instituto de F\'{\i}sica,
           Universidade Federal do Rio Grande do Sul, PO Box 15051,
           91501--970 Porto Alegre, RS, Brazil\\
  \inst{3} Grupo de Sistemas Complejos and Departamento de F\'{\i}sica,
           Escuela T\'ecnica Superior de Ingenieros Agr\'onomos,
           Universidad Polit\'ecnica de Madrid, 28040 Madrid, Spain\\
  \inst{4} Instituto Mixto de Ciencias Matem\'aticas CSIC--UAM--UC3M-UCM,
           Universidad Aut\'onoma de Madrid, Cantoblanco, 28049 Madrid, Spain
}

\abstract{
We demonstrate the existence of superscarring in the LiNC$\leftrightharpoons$LiCN
isomerization reaction described by a realistic potential interaction
in the range of readily attainable experimental energies.
{This phenomenon arises as}
the effect of two periodic orbits appearing ``out of the blue''
in a saddle--node bifurcation taking place in the dynamics of the system.
Potential practical consequences of
{this superlocalization}
in the corresponding wave functions are also considered.
}

\begin{document}

\maketitle

\section{Introduction}
The localization of certain quantum eigenstates along unstable
periodic orbits (PO) in classically chaotic
systems -- \emph{i.e.}, \emph{quantum scarring} \cite{Heller1} --
has profound consequences for the dynamics of quantum systems.
It was first observed in numerical calculations on the Bunimovitch
stadium billiard \cite{McDonald}.
At the most basic level, scarring \cite{Bogomolny,scars} is a spectacular
example of the influence of classical structures,
\emph{i.e.}\ PO \cite{KaplanHeller} and their associated
manifolds \cite{Tomsovic,Vergini2,Homo} on quantum behavior \cite{Gutzwiller}.
Scarring is responsible, for example, for significant deviations
from the predictions of random matrix theory \cite{Kaplan}.
It has also been conjectured recently that they are important
in open systems to understand the semiclassical structure
of resonances \cite{Novaes,Saraceno}, the number of which {has been seen to scale}
with $\hbar$ according to a fractal Weyl law
\cite{fractal_Weyl,Ramilowski}.
At the same time, scarring has important practical
consequences in various fields, including
tunnel currents in mesoscopic devices \cite{nano},
quantum dots \cite{qdots},
optical cavities \cite{optcav},
optical fibers \cite{optfib}, and very recently the existence
of relativistic quantum scars has been demonstrated \cite{Huang}
in the emerging field of graphene applications \cite{Guinea}.

The vast majority of the work published on scars has concentrated
on fully chaotic systems, where the POs are isolated and unstable.
However, Keating and Prado (KP) \cite{Keating} recently considered
the specific peculiarities of generic systems (with mixed dynamics),
where bifurcations of the scarring PO occur.
By extending Bogomolny theory \cite{Bogomolny}, they found that the
density localization around bifurcated POs are wider and with
amplitudes at least as large as those far from bifurcations.
This led KP to introduce the term \emph{superscars} {to name}
this new localization phenomenon which they illustrated with
numerical examples in perturbed cat maps.
The influence of bifurcating orbits on the density of states
was studied by Schomerus  and Sieber \cite{SS}, who also
found different scaling exponents on the dependence of such
magnitude with $\hbar$.
Due to the nature of the arguments used by KP,
which only apply to high excitation energies
($\hbar \rightarrow 0$ limit),
an open question has remained and concerns the transferability
of KPs conclusions beyond maps to actual physical systems over
experimentally realizable energy ranges.

In this Letter, we address this crucial point, namely the existence
of a universal behavior in the exponents of the moment distribution
characterizing superscars \cite{Keating}.
This is done in the context of a realistic model for the
LiNC$\leftrightharpoons$LiCN isomerization reaction \cite{LiCN}.
More specifically, for this system we investigate the superscarring
associated with POs which originate in a saddle--node (SN)
bifurcation of the isomerization dynamics \cite{Zembekov}.

Isomerization plays a fundamental role in many biochemical
processes and it is an ideal benchmark for dynamical systems
theory \cite{LiCN,Zembekov}.
Also, the example that we have chosen can be considered as a
prototype for this kind of processes (see description of the model below),
which has the added bonus of exhibiting a rich and complex vibrational dynamics.
Moreover, the superscarring phenomenon discussed here is closely
related to some particular aspects of the transition state (TS)
theory (TST) \cite{TST} at quantum level.
This theory, that was developed in the 1930's after the seminal
work of Eyring \cite{Eyring} and Wigner \cite{Wigner},
envisages the physical processes associated to chemical
reactions as ``skiing the reaction slopes'' \cite{Marcus}
or landscapes \cite{landscapes}, so that the barriers existing
between reactants and products are surpassed.
Quantum mechanically, these barriers accommodate the corresponding
Gamow--Siegert resonant states \cite{Seideman,Skodje},
and the topology of the associated wave functions
{configures}
the doorways controlling the quantum reactivity \cite{TST2}.

In the title reaction there exists, apart from the reaction barrier
corresponding to the saddle point in the potential energy surface
that was previously studied in Ref.~\cite{Llorente},
another interesting barrier of dynamical origin \cite{Pollak,Gaspar},
associated to the POs arising from a SN bifurcation in the
corresponding dynamics \cite{Zembekov}.
According to KP's results the characteristics of this barrier
changes dramatically with the excitation energy.
At high energy, where the two POs are isolated the quantum density
localization effect due to scarring is less pronounced.
On the contrary, when lower values of the energy,
close to the bifurcation point, are considered a strong
superlocalization  takes place.
The fundamental difference in the widths taking place in this
two different situations is expected to have some relevant
influence in the corresponding flux \cite{wp},
leading to a different behavior in the respective reactive dynamics.

Finally, it should be pointed out that other interesting,
although completely different, effects of SN bifurcations
in the vibrational dynamics of molecular systems (HCP molecule)
have been described in the literature \cite{HCP1}.

\section{Model and calculations}

As a working model, we choose the LiNC/LiCN isomerizing
molecular system, which has been extensively studied in connection
with quantum chaos \cite{LiCN}.
This system is representative of a large class of small polyatomic
molecules, which exhibit similar behavior, mainly due to the presence
of a large amplitude (floppy) motion in one of the vibrational modes.
This class includes other cyanides, such as HCN/HNC \cite{HCN}),
alkaline cyanides \cite{alkalineCN}, and other similar especies, such as,
HCP \cite{HCP1}, the \chem{HO_2} radical, or the van der Waals complexes.

\begin{figure}
 \onefigure[width=3.0cm]{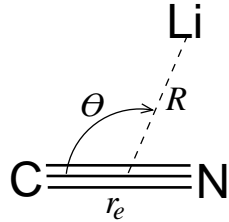}
 \caption{Definition of the Jacobi coordinates used in our calculations.}
 \label{fig:0}
\end{figure}

\begin{figure}
 \onefigure[width=8.0cm]{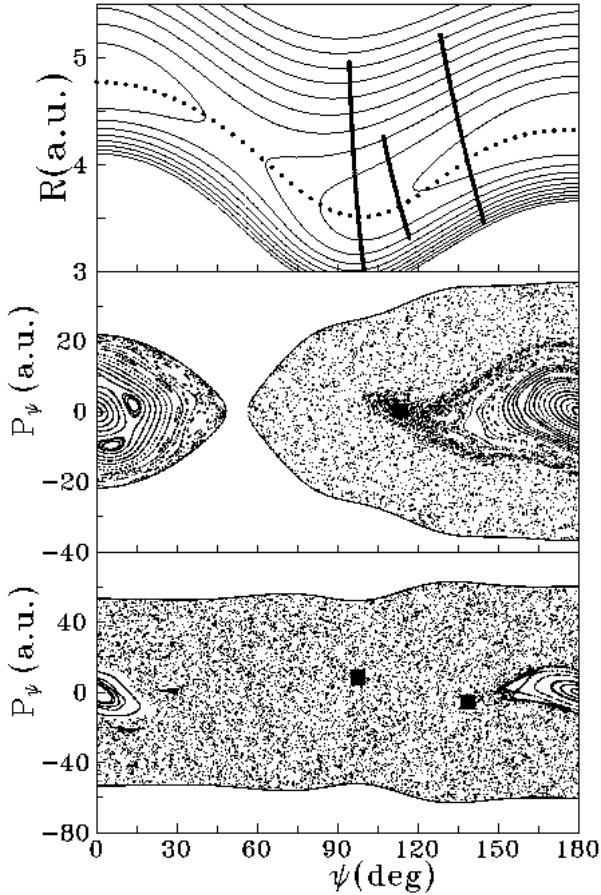}
 \caption{(Top) Potential energy surface for LiNC/LiCN.
   The minimum energy path (dotted line), and three saddle--node
   periodic orbits (full line) at $E_\mathrm{bif}$=3440.64 cm$^{-1}$
   (central orbit) and $E_1$=9196.0 cm$^{-1}$ (lateral trajectories)
   relevant to our work have been plotted superimposed.
   (Middle and bottom) Composite Poincar\'e surface of sections for
   the same energies.
   The position of the fixed points corresponding to the
   periodic orbits are represented with full squares.}
 \label{fig:1}
\end{figure}

Our system can be modelled by a realistic two degrees of freedom
vibrational ($J=0$) Hamiltonian, which is given by
%
\begin{equation}
  H=\frac{P_R^2}{2\mu_1}+
    \frac{1}{2}\left(\frac{1}{\mu_1 R^2} +
    \frac{1}{\mu_2 r_e^2} \right) P_\vartheta^2 + V(R,\vartheta).
  \label{eq:1}
\end{equation}
Here we use Jacobi coordinates, which are common for example in scattering theory
\cite{ScatteringTh} and celestial mechanics \cite{CelestialM}.
In our case these coordinates, ($r,R,\vartheta$), are defined as the C--N distance,
the distance between the Li atom and the center of mass of the CN fragment,
and the angle formed by these two vectors, respectively
{(see Fig.~\ref{fig:0}).}
%
The corresponding reduced masses are defined as
$\mu_1=m_\mathrm{Li} m_\mathrm{CN}/m_\mathrm{LiCN}$ and
$\mu_2=m_\mathrm{C} m_\mathrm{N}/m_\mathrm{CN}$.
The C--N motion is kept frozen at its equilibrium value, $r_e$,
since the associated frequency is high, thus decoupling very effectively
from the rest of coordinates in the molecule.
The interaction potential is expressed as a series {of}
Legendre polynomials,
%
\begin{equation}
  V(R,\vartheta) = \sum_{\lambda=0}^9 P_\lambda(\cos \vartheta) v_\lambda(R),
\end{equation}
whose coefficients consist of a sum of short-range and long-range
{contributions whose explicit forms}
have been taken from the literature \cite{Essers}.
It is shown in the top panel of Fig.~\ref{fig:1} as a contours plot.
%
As can be seen, it presents two wells at $\vartheta=0$ and 180$^\circ$,
corresponding to the linear isomers LiCN and LiNC, which are separated
by a modest energy barrier.
The motion in $\vartheta$ is very floppy
and chaos sets in at moderate values of the excitation energy.
Superimposed to the potential surface, we also present
 {in the plot the POs
which are relevant for our discussion on superscars,
at two different values of the interaction energy.}
They all run almost vertically, and correspond to a 1:1 resonance
which appear ``out of the blue'' due to a SN or tangent bifurcation
at $E_\mathrm{bif}=3440.64$ cm$^{-1}$, as discussed in Ref.~\cite{Zembekov}.
At this energy, the two orbits are indistinguishable,
since they both coalesce in the central (shorter) trajectory shown
in the figure.
As energy increases the two POs separate, the unstable one moving
to the right and the stable towards the left.
This stability is later lost for $E>4960.1$ cm$^{-1}$.
These two POs have also been plotted in the top panel of Fig.~\ref{fig:1}
for an energy far away from the bifurcation point,
namely $E_1 = 9196.0$ cm$^{-1} \simeq 3 E_\mathrm{bif}$.

The corresponding classical dynamics can be adequately followed by computing
Poincar\'e surfaces of section (SOS) using the minimum energy path,
$R_e(\vartheta)$, connecting the two isomers, as the sectioning plane.
This choice requires {an extra} canonical transformation,
%
\begin{eqnarray}
 \psi   & = & \vartheta, \nonumber \\
 P_\psi & = & P_\vartheta-
   \left(\frac{\upd R_e(\vartheta)}{\upd\vartheta}\right)_{\vartheta=\psi} P_R,
\end{eqnarray}
to make the SOS an area preserving map \cite{Ezra}.
Composite SOSs for $E_\mathrm{bif}$ and $E_1$ are shown in the
middle and bottom panels of Fig.~\ref{fig:1}, respectively.
Two comments are in order.
First, the dynamics is mixed, being chaotic in the region of interest --
\emph{i.e.} where the two bifurcated POs sit -- {and} regular
around the isomer configurations.
Second, the obvious signature of a cantori \cite{Losada},
seen as an accumulation of points {next to} the region of stability
{around} $\vartheta$=180$^\circ$, is apparent
for $E_\mathrm{bif}$, and this bottleneck disappears at $E_1$.

As mentioned before, one critical element in our work is the
computation of the scarred wave functions.
The reason for this should be clear.
Since we want to perform our study far from the high $E$ (low $\hbar$)
semiclassical limit,
where the quantum oscillations in the local configurations and energy
averages implied in scarring \cite{Bogomolny} easily cancel out,
an algorithm able to introduce the subtle dynamical
 {effects}
associated to the scarring PO in the scar functions is forcefully needed.
In our case, this is achieved by using
the method described in \cite{Polavieja}
consisting in time evolving a wave packet, $|\phi(t)\rangle$,
initially launched at one of the turning points of the PO under study,
obtaining afterwards all the relevant information by Fourier transform.

\begin{figure}
 \onefigure[width=8.0cm]{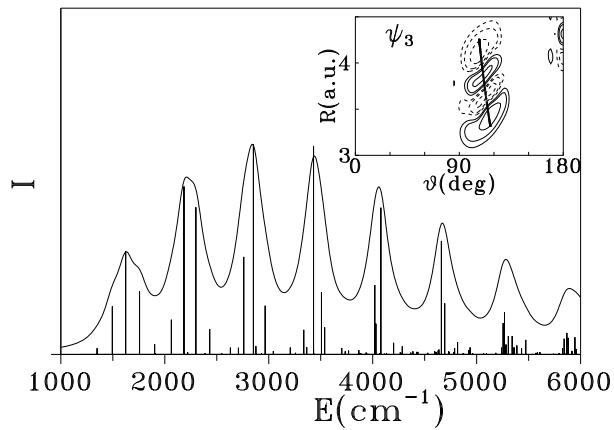}
 \caption{Infinite resolution spectrum and its low
   resolution version generated from a wave packet localized
   along the central 1:1 periodic orbit on the top panel
   of Fig.~\protect\ref{fig:1}.
   The inset shows the scar function $\psi_3$, whose mean energy
   roughly correspond to $E_\mathrm{bif}$.}
 \label{fig:2}
\end{figure}

We start with the (infinite resolution) stick spectrum:
%
\begin{eqnarray}
    I_\infty(E) & = & 
      \int_{-\infty}^{+\infty}
      \upd t \; \mathrm{e}^{\mathrm{i}Et/\hbar} \langle \phi(0)|\phi(t) \rangle \nonumber \\
     & = & \sum |\langle\phi(0)|n\rangle|^2 \delta(E-E_n),
 \label{eq:2}
\end{eqnarray}
where kets $|n\rangle$ {represent} the eigenfunctions of the system.
An example of the results obtained from such calculations is shown
in Fig.~\ref{fig:2}.
As can be observed, the distribution of sticks, both in separation and length,
is rather irregular, but when examined closely they are seen to
{be} nicely grouped in clumps.
This regularity is due to {the} recurrences in the autocorrelation
function induced by the PO motion \cite{KaplanHeller}, {and it} shows up as
well--defined bands in the low resolution version of the spectrum,
%
\begin{equation}
  I_T(E)=
       \int_{-\infty}^\infty \upd t \;
       \mathrm{e}^{-t^2/T^2}
       \mathrm{e}^{\mathrm{i}Et/\hbar} \langle \phi(0)|\phi(t) \rangle.
  \label{eq:3}
\end{equation}
where the cutoff parameter $T$ has been introduced in order to eliminate
the long term dynamics not directly associated to the scarring process.
The result for $T$ equal to the period of the scarring PO is
displayed in Fig.~\ref{fig:2}.
Here, a series of bands are seen to appear centered at the discrete
values of the energy, $E^\mathrm{BS}_n$, that {are} obtained semiclassically
using {the} suitable Bohr--Sommerfeld (BS) quantization condition of the
action along the PO [see eq.~(\ref{eq:5}) below].
In a second step, we focus our attention in one single of these bands,
and calculate the scar wave {function}, $|\psi_n\rangle$, associated
to it by Fourier transforming the original packet, $|\phi(t)\rangle$,
at the corresponding BS energy, $E^\mathrm{BS}_n$,
%
\begin{equation}
   |\psi_n\rangle = 
     \int_{-\infty}^{\infty} \; \upd t \; \mathrm{e}^{-t^2/T_E^2} \;
     \mathrm{e}^{\mathrm{i}(E^\mathrm{BS}_n-\hat{H})t/\hbar} |\phi(0)\rangle.
  \label{eq:4}
\end{equation}
Here we choose the Ehrenfest time \cite{footnote2} as the
value for the cutoff parameter $T$, since it is known \cite{Vergini2}
that this procedure guarantees that the dynamical information
concerning the hyperbolic structure associated to the scarring PO
is embedded in the topology of $|\psi_n\rangle$.
%
By expanding $| \phi(0) \rangle$ in the basis set of the system eigenstates,
the integral in eq.~(\ref{eq:4}) can be solved explicitly, and one obtains
%
\begin{equation}
  |\psi_n \rangle =\sqrt{\pi} T_E  \sum_n  \mathrm{e}^{-(E_n-E_{BS})^2 T^2/4 \hbar^2}
     \langle n |  \phi(0) \rangle        |n \rangle.
\end{equation}
{This equation clearly shows how the scarred states defined by us
fulfill the two most important criteria accepted in scar theory.
On the one hand, they are built by averaging over the eigenstates
within an energy window of width $\hbar /T_E$ around $E_n^{\rm BS}$,
in the spirit of Bogomolny's interpretation of scarring \cite{Bogomolny}.
On the other hand, the particular average that we use incorporates the dynamical
correlations existing in $ \langle \phi(0) |  \phi(t) \rangle $,
in agreement with the view introduced by Heller \cite{KaplanHeller}.}
It should be stressed that these correlations are so strong that the
scar functions acquire a very pronounced semiclassical character,
independently of particular details {of} the individual eigenfunctions
used {in} the construction,
{and in this sense our method is very robust.}
In the inset to Fig.~\ref{fig:2} we show as an example $|\psi_3\rangle$,
whose (mean) energy roughly corresponds to $E_\mathrm{bif}$
and it is constructed with the eigenstates in the fourth clump.
 {Let us imagine} now that $\hbar$ were reduced to one half {of its} value.
The energy density would increase, and the clumps of Fig.~\ref{fig:2} would {shift} to the left,
i.e.~to the region of smaller energies.
The eigenfunctions contributing to {each} clump would then be different,
but {our computed scar wave functions would be the same.}

In conclusion, a scar function is {only} characterized by
{the} number of excitations along the PO,
and this number depends both on the
{values of $E$ and $\hbar$, which are given by}
the Bohr-Sommerfeld quantization condition.
{Another}
relevant characteristic of the scar functions is related to the direction transverse to the PO.
{In the case of an unstable PO}, a hyperbolic Hamiltonian dominates
the {flux in its neighborhood} and the wave packet, $|\phi(0)\rangle$,
moves quickly along the {associated} manifolds.
On the other hand, at bifurcations the dynamics acquire some stability,
and the original wave packet move away from the central PO, but in a {more} slowly way.
These two different behaviors {have an influence} in the width of the
{related} scar functions, {as will be demonstrated in the present} article.

\section{Results and discussions}
\begin{figure*}
\begin{center}
 \onefigure[width=11cm]{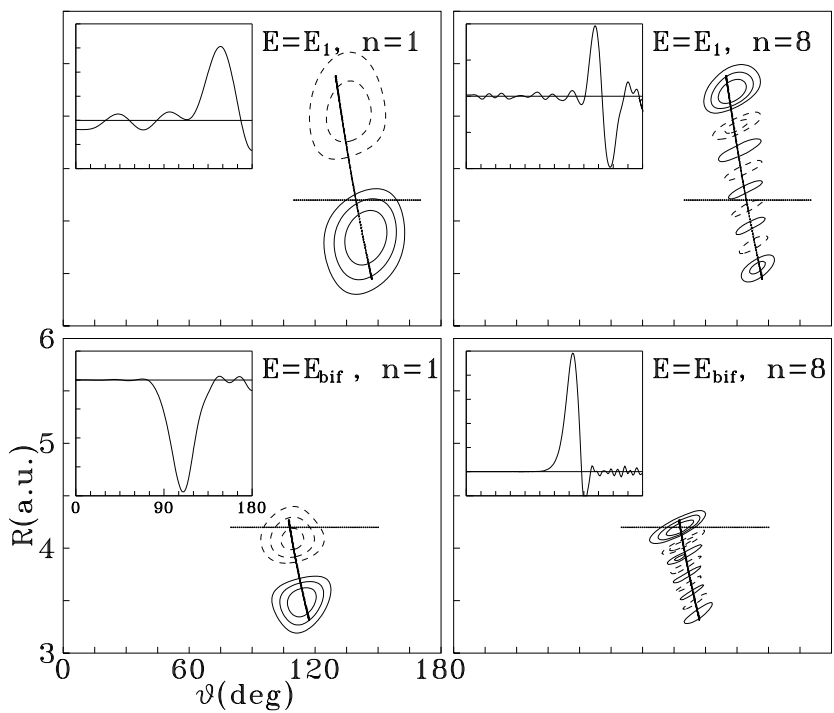}
 \caption{Some scar functions computed with Eq.~(\ref{eq:4})
  at different values of $\hbar$ obtained from Eq.~(\ref{eq:3}).
  The scarring periodic orbit is superimposed.
  The insets show the wave function profiles along one of the
  cuts at constant $R$ (shown as short horizontal segments)
  which are used in Eq.~(\ref{eq:6}).}
 \label{fig:3}
\end{center}
\end{figure*}

Let us next quantify the scarring effect on $\psi_n$ due to the
1:1 SN POs that we are considering.
As described by KP \cite{Keating}, this can be achieved with the aid
of the second moment of the wave function amplitude distribution
averaged over regions shrinking as $\hbar \rightarrow 0$ but
containing an increasing number of de Broglie wavelengths.
In the case of an isolated unstable PO this magnitude follows,
in the limit, a universal power law $\hbar^{1/2}$.
{However,} for just bifurcated POs of the type that we are considering here,
{i.e.}~tangent bifurcation, this expression goes as $\hbar^{1/3}$
due to the superscarring effect of the bifurcated orbits\cite{Keating}.
In order to carry out this calculation, we first select two values of
the vibrational energy, one corresponding to the bifurcation energy,
$E_\mathrm{bif}$, and the other one, $E_1$, sufficiently
far away from it to ensure that the PO is isolated.
Second, we compute a series of scar functions, $|\psi_n\rangle$,
with different values of the excitation quantum number $n$ at
the same energy, $E$.
In order to do this, we vary $\hbar$ according to the BS quantization
condition for the PO
%
\begin{equation}
   \hbar(n)=S_E \left(n+\frac{\nu_E}{4}\right)^{-1},
  \label{eq:5}
\end{equation}
{where} $S_{E_\mathrm{bif}}$=3.5, $\nu_{E_\mathrm{bif}}=2$,
$S_{E_1}$=13.75, $\nu_{E_1}=3$ in our case.
Notice that for the second energy the Maslov index has increased
by one due to the appearance of a self--focal point in the PO.
The net effect of the $\hbar$ scaling represented by (\ref{eq:5})
is to shift the center of the {$n$--th} band to the
value of the working energy, $E_\mathrm{bif}$ or $E_1$.
Some scar wave functions computed in this way are shown in Fig.~\ref{fig:3}.
%

\begin{figure}
 \onefigure[width=8.0cm]{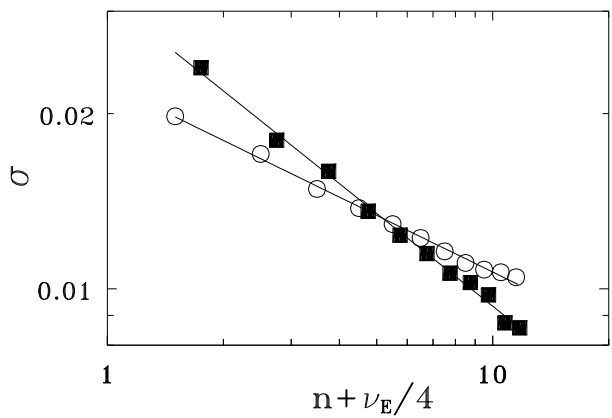}
 \caption{Log--log plot of the averaged scar wave functions widths,
   computed by Eq.~(\ref{eq:7}), \emph{vs} ($n+\nu_E/4$),
   for two values of the vibrational energy:
   $E_\mathrm{bif}=3440.64$ cm${-1}$ (open circles)
   and $E_1=9196.0$ cm${-1}$ (full squares).
   Also shown are the least squares fitted straight lines corresponding
   to power laws, $(n+\nu_E/4)^{-\alpha}$ with scaling exponents
   0.32$\pm$0.01 and 0.53$\pm$0.02, respectively, indicating that we are
   in the presence of superscarrings.}
 \label{fig:4}
\end{figure}

As can be seen, they all correspond to different excitations in the $R$ motion
{with} no nodes in the $\vartheta$ coordinate.
Moreover, their widths (transverse to the PO) change very slowly at $E_\mathrm{bif}$,
which is a distinctive property of the phenomenon of superscarring.
In order to make this assertion quantitative, we evaluate the variance
of the probability density at a fixed value of $R$,
%
\begin{equation}
  \sigma_R^2(n) = \int \upd \vartheta \; \vartheta^2 \frac{|\psi_n(R,\vartheta)|^2}{W(R)} -
  \left[\int \upd \vartheta \; \vartheta \frac{|\psi_n(R,\vartheta)|^2}{W(R)} \right]^2,
  \label{eq:6}
\end{equation}
with $W(R)=\int d\vartheta \; |\psi_n(R,\vartheta)|^2$.
To illustrate this procedure, the profile of each function in
Fig.~\ref{fig:3} is shown for $R=4.2$ a.u.\ in the insets.
Then, we calculate the mean width weighted by $W(R)$ as
%
\begin{equation}
  \sigma(n) = \left[ \int \upd R \; W(R) \sigma_R^2(n) \right]^{1/2}.
  \label{eq:7}
\end{equation}
The corresponding results for $E_\mathrm{bif}$ and $E_1$ are shown
in Fig.~\ref{fig:4}.
Several comments are in order.
First, as can be seen in the log--log plot both sets of data follows
remarkably well a scaling power law, $(n+\nu_E/4)^{-\alpha}$,
with the excitation quantum number, $n$ (equivalent to $\hbar$).
Second, from the slopes of least squares fittings we have that
$\alpha_{E_\mathrm{bif}}=0.32\pm 0.01$ and $\alpha_{E_1}=0.53\pm0.02$.
Third, these values agree extremely well, within the numerical error,
with the universal behavior predicted in the theory of KP,
according to which the exponent is 1/3 at the bifurcation point,
and 1/2 far from the bifurcation, respectively.

This last result confirm beyond any doubt that the SN POs studied
in this Letter, which corresponds to an excited molecular motion
for the LiNC/LiCN isomerizing molecular systems taking place mainly
along the stretching coordinate, $R$, is able to produce superscarring
on the corresponding quantum wave functions for energies near the
bifurcation point, and this effect disappears as we move away form
this point.
%

\section{Conclusions}
Summarizing, in this Letter we have reported for the first time
conclusive evidence for the existence of superscarring associated
to bifurcated POs \cite{Keating} in a realistic molecular system.
More interestingly, this happens in the regime of very low
excitations along the PO, which opens the possibility for
considering the influence of scarring in quantum processes where
the probability flux is strongly affected by the existence of
dynamical barriers.
In particular, the effect of scarring on LiNC$\leftrightharpoons$LiCN
isomerization rates and specially its dependence on the value of the
vibrational excitation energy is a topic that is currently under investigation.


\acknowledgments
This work has been supported by the Ministerio de Ciencia e Innovaci\'on
(Spain) under projects MTM2006--15533,
MTM2009-14621 and CONSOLIDER 2006--32 (i--Math),
Comunidad de Madrid under the project S--0505/ESP--0158 (SIMUMAT),
and a grant (to SDP) from Fundaci\'on Carolina (AECI--Spain).


\begin{thebibliography}{99}
\bibitem{footnote2} The Ehrenfest time is defined as $T_E=(2\lambda)^{-1}\ln (S/\hbar)$,
$\lambda$ being the stability index of the PO and $S$ a typical action transverse
to it (here taken as the area of the surface of section).
\end{thebibliography}
\end{document}